\documentclass[%
 reprint,
superscriptaddress,
 amsmath,amssymb,
 aps,prl,
]{revtex4-2}

\usepackage{graphicx}
\usepackage{dcolumn}
\usepackage{bm}
\usepackage{hyperref}
\usepackage{latexsym}
\usepackage{amsmath, amsthm, amssymb}
\usepackage{relsize}
\usepackage{epsfig}
\usepackage{multirow}
\usepackage{array}
\usepackage[dvipsnames]{xcolor}
\usepackage{placeins}
\usepackage[normalem]{ulem}
\begin{document}

\preprint{APS/123-QED}

\title{Phase separation kinetics of 2-TIPS at low density: Cluster growth by ballistic agglomeration}

\author{Nayana Venkatareddy\textsuperscript{\textdaggerdbl}}
 \affiliation{Department of Physics, Indian Institute of Science, C. V. Raman Ave, Bengaluru 560012, India}
 \thanks{These authors contributed equally to this work.}
\author{Partha Sarathi Mondal\textsuperscript{\textdaggerdbl}}%
\affiliation{Department of Physics, Indian Institute of Technology (BHU) Varanasi, Uttar Pradesh 221005, India}%
\thanks{These authors contributed equally to this work.}
\author{Shradha Mishra}
\email{smishra.phy@itbhu.ac.in}
\affiliation{Department of Physics, Indian Institute of Technology (BHU) Varanasi, Uttar Pradesh 221005, India}
\author{Prabal K. Maiti}
\email{maiti@iisc.ac.in}
\affiliation{Department of Physics, Indian Institute of Science, C. V. Raman Ave, Bengaluru 560012, India}

\date{\today}

\begin{abstract}
We study the kinetics of two-temperature induced phase separation (2-TIPS) in dilute binary mixtures of active (“hot”) and passive (“cold”) particles using molecular dynamics simulations and a coarse-grained hydrodynamic model. Following a temperature quench, cold particles nucleate into mobile clusters that move ballistically and merge through successive coalescence events. The resulting domain growth exhibits dynamic scaling with a growth exponent \(\approx0.7\), markedly faster than diffusive coarsening. We identify this regime as ballistic agglomeration of cold clusters, demonstrating a distinct nonequilibrium growth mechanism in low-density scalar active systems.
\end{abstract}

\maketitle


The study of phase separation kinetics constitutes a central theme in condensed matter physics, elucidating universal features such as scaling behavior, domain growth, and morphological self-similarity during coarsening \cite{puri2009kinetics,bray1994theory,hohenberg1977theory,cates2018theories}. Beyond the theoretical importance, understanding the kinetics is important for many practical applications, for instance, in the formation of micro-structures during the manufacturing of materials \cite{porter2009phase}. In recent years, with the advances in active matter systems \cite{marchetti2013hydrodynamics,romanczuk2012active,toner2005hydrodynamics,ramaswamy2019active}, the importance of kinetic processes has garnered new interest in the context of far from equilibrium systems showing Motility-Induced Phase Separation (MIPS) \cite{annurev:/content/journals/10.1146/annurev-earth-063016-020226}, swarming and flocking \cite{VICSEK201271,toner2005hydrodynamics}, and active turbulence \cite{doi:10.1073/pnas.1202032109}.

In conventional active matter systems, the activity emerges through internal force generation mechanisms of individual agents. In contrast to these systems, in a binary mixture, activity can manifest through effective temperature difference resulting from the difference in the dynamical characteristics of the constituent species - referred to as hot and cold \cite{weber2016binary,chari2019scalar}. When the temperature difference exceeds a density dependent critical value, the hot and cold particles phase separate, and is termed two temperature induced phase separation (2-TIPS). The two-temperature framework has been effectively applied to diverse biological systems like chromosomal dynamics \cite{ganai2014chromosome}, behavior of molecular motors \cite{li2017double}, enzymes \cite{PhysRevLett.123.128101}, and cell motility–driven demixing \cite{mccarthy2024demixing}. Additionally, the 2-TIPS framework can serve as a minimal model for tracer dynamics in nonequilibrium media \cite{al2025statistical}.

Owing to the versatility of the two temperature framework, 2-TIPS phenomena have been studied extensively \cite{chari2019scalar,PhysRevE.107.034607,PhysRevE.107.024701,D3SM00796K,chattopadhyay2021heating,chattopadhyay2024stability,grosberg2018dissipation,grosberg2015nonequilibrium,PhysRevResearch.2.023200}, wherein the cold particles spontaneously assemble into dense clusters exhibiting crystalline order in Lennard-Jones (LJ) monomers and liquid-crystalline (LC) order in spherocylinders. Despite substantial progress in understanding the steady-state behavior, the kinetics of 2-TIPS have remained comparatively unexplored. Recently, we demonstrated that for LJ monomers, the domain growth during 2-TIPS follows a $\sim t^{1/3}$ scaling, both in two and three dimensions, in the high-density regime \cite{1c8b-hmxv}, reminiscent of the Lifshitz–Slyozov (LS) growth law in equilibrium binary mixtures (EBMs)\cite{LIFSHITZ196135}. In EBMs, both the mechanism and kinetics of phase separation are known to depend sensitively on density \cite{bray1994theory,tanaka2000viscoelastic}, a feature also observed in active systems \cite{fily2014freezing,pattanayak2021ordering}. It is natural to ask whether two-temperature systems display a similar behavior.

In this Letter, we study the kinetics of 2-TIPS in a low-density binary mixture of hot and cold LJ particles using molecular dynamics (MD) simulations and a corresponding coarse-grained model (CG) coupled to hydrodynamic flow in two-dimensions (2\(d\)). We find that phase separation begins with the nucleation of cold clusters, which subsequently move ballistically and grow via coalescence, resulting in significantly faster domain growth. The growth exponent obtained from MD simulations is  $\approx 0.70$, which is in good agreement with the predictions of ballistic aggregation theory and the results obtained from the CG model.

{\em Molecular dynamics (MD):} We first discuss the model and results obtained from the  MD simulations in the NVT ensemble on a symmetric binary mixture consisting of 80,000 hot and cold particles in a 2\(d\) periodic box. All the particles interact with each other through the LJ potential described below.
 \begin{equation}
    U_{LJ}(r)=4\epsilon\bigg[\bigg(\frac{\sigma}{r}\bigg)^{12}-\bigg(\frac{\sigma}{r}\bigg)^6\bigg]
\end{equation}
where \(\epsilon\) denotes the strength of the interaction, \(\sigma\) is the diameter of the LJ particle, and \(r\) is the distance between the interacting particles. The results reported here are presented in reduced units, where the length and energy are expressed in units of \(\sigma\) and \(\epsilon\), respectively.  The simulations in the present work are carried out at the low density of \(\rho^*=0.1\), and the Nos\'e Hoover \cite{evans1985nose} thermostat is used to maintain the temperatures of hot (\(T_h^*\)) and cold particles (\(T_c^*\)). Starting from a well-mixed state with \(T_c^*=T_h^*=2\), the system is quenched into the phase-separated regime by instantaneously raising \(T_h^*\) to the desired value and then evolved for 10 million (M) time steps. Additional details on the simulation are provided in Section 1 of the Supplemental Material (SM).

\begin{figure}[t]
\centering
\includegraphics[scale=0.025]{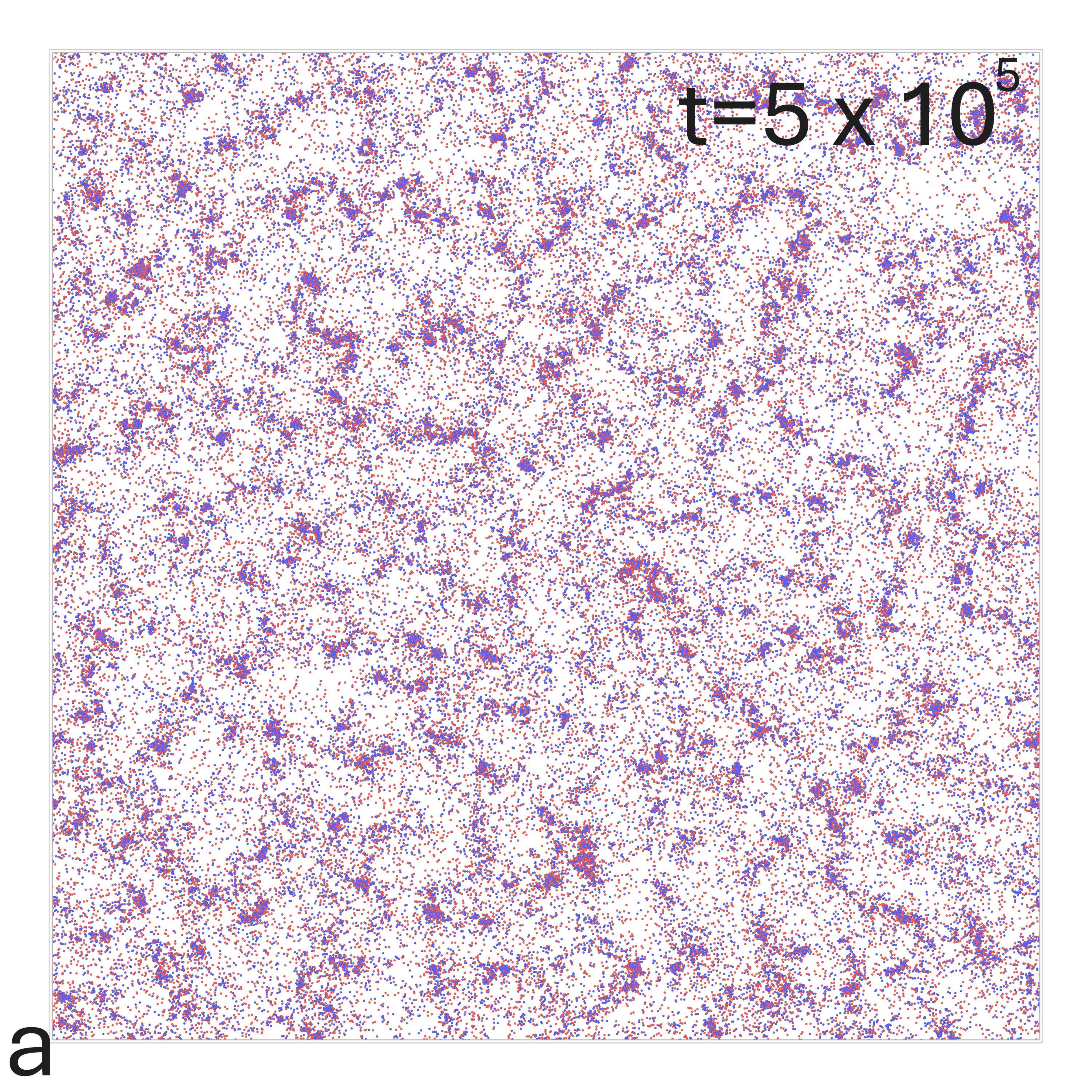} 
\includegraphics[scale=0.025]{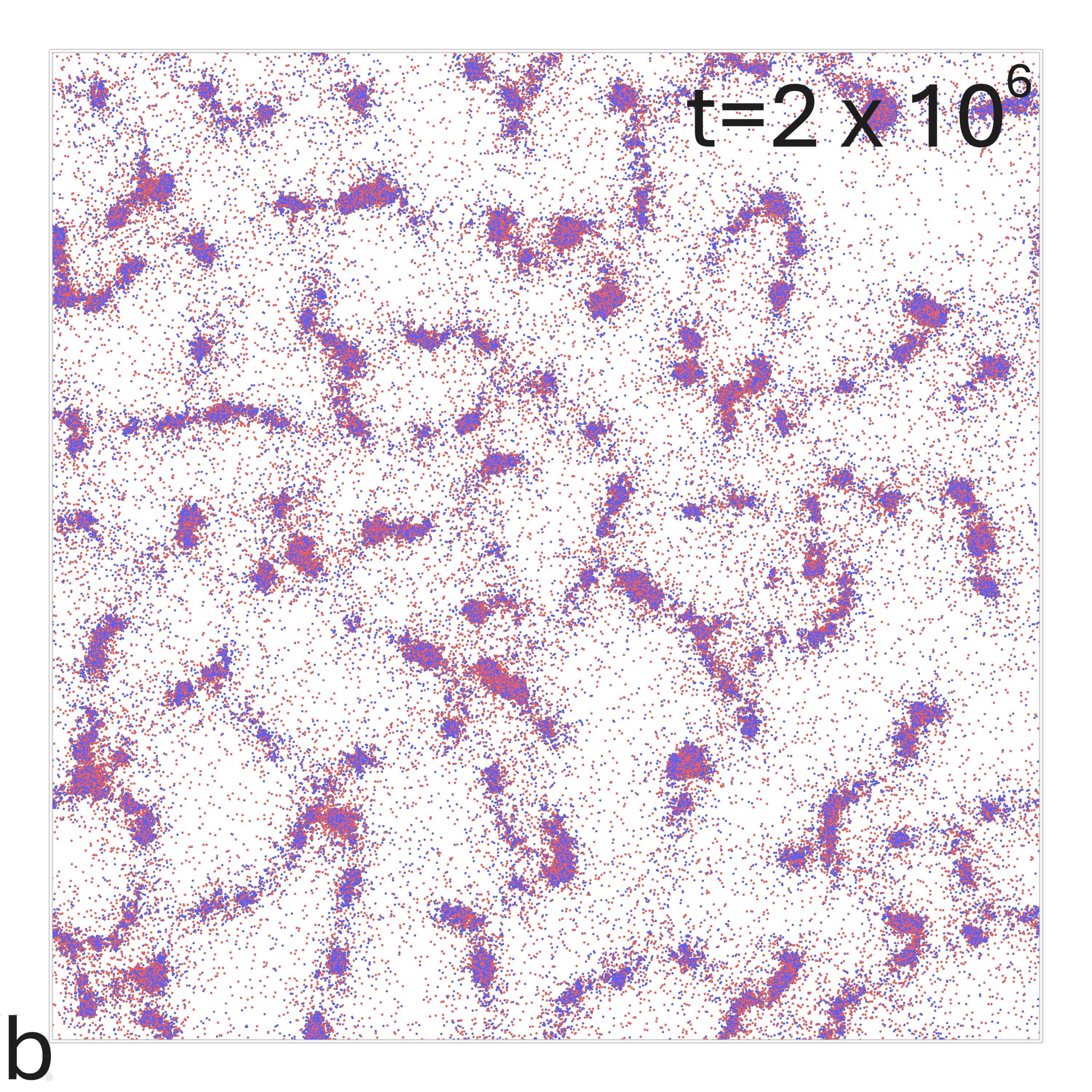}
\includegraphics[scale=0.025]{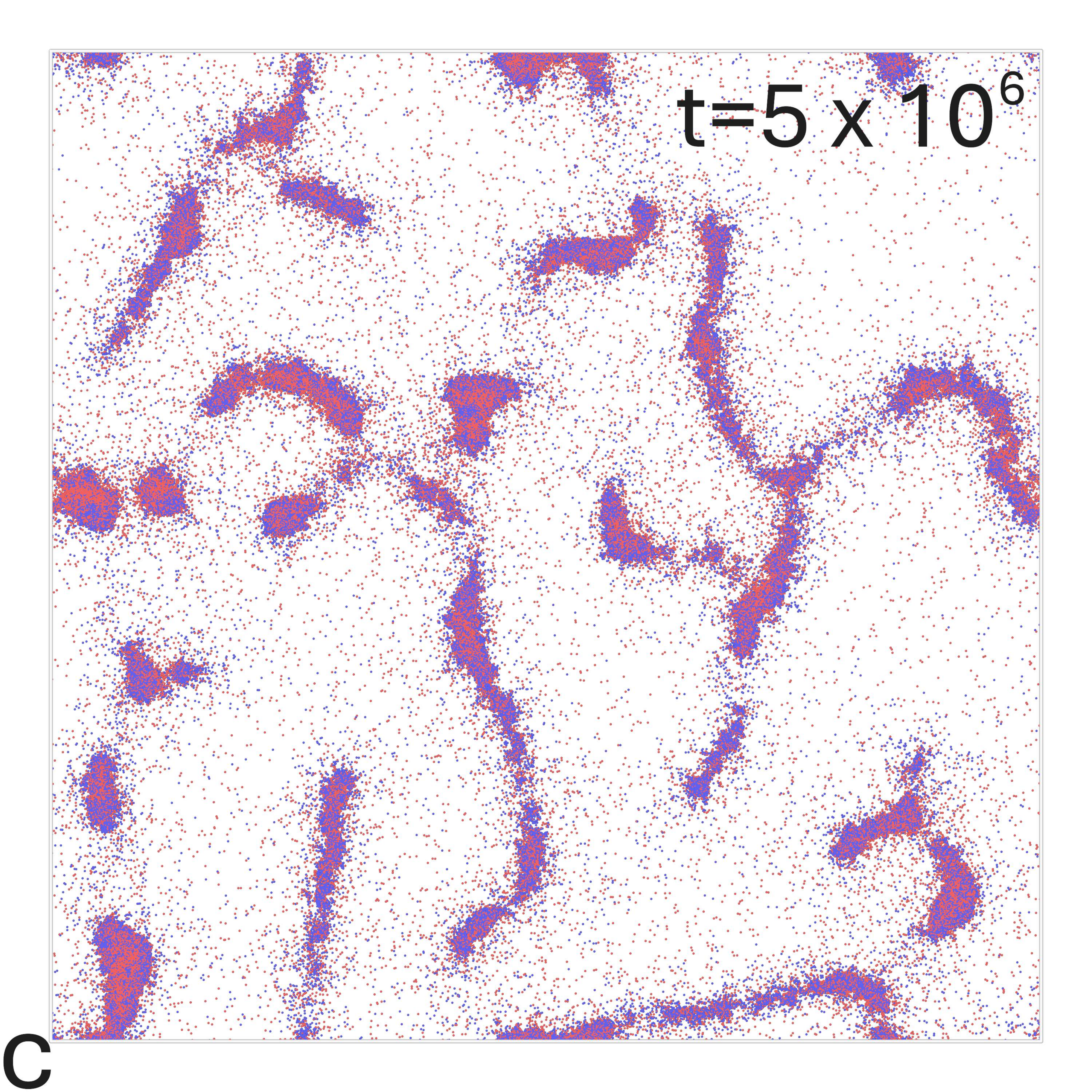} 
\includegraphics[scale=0.105]{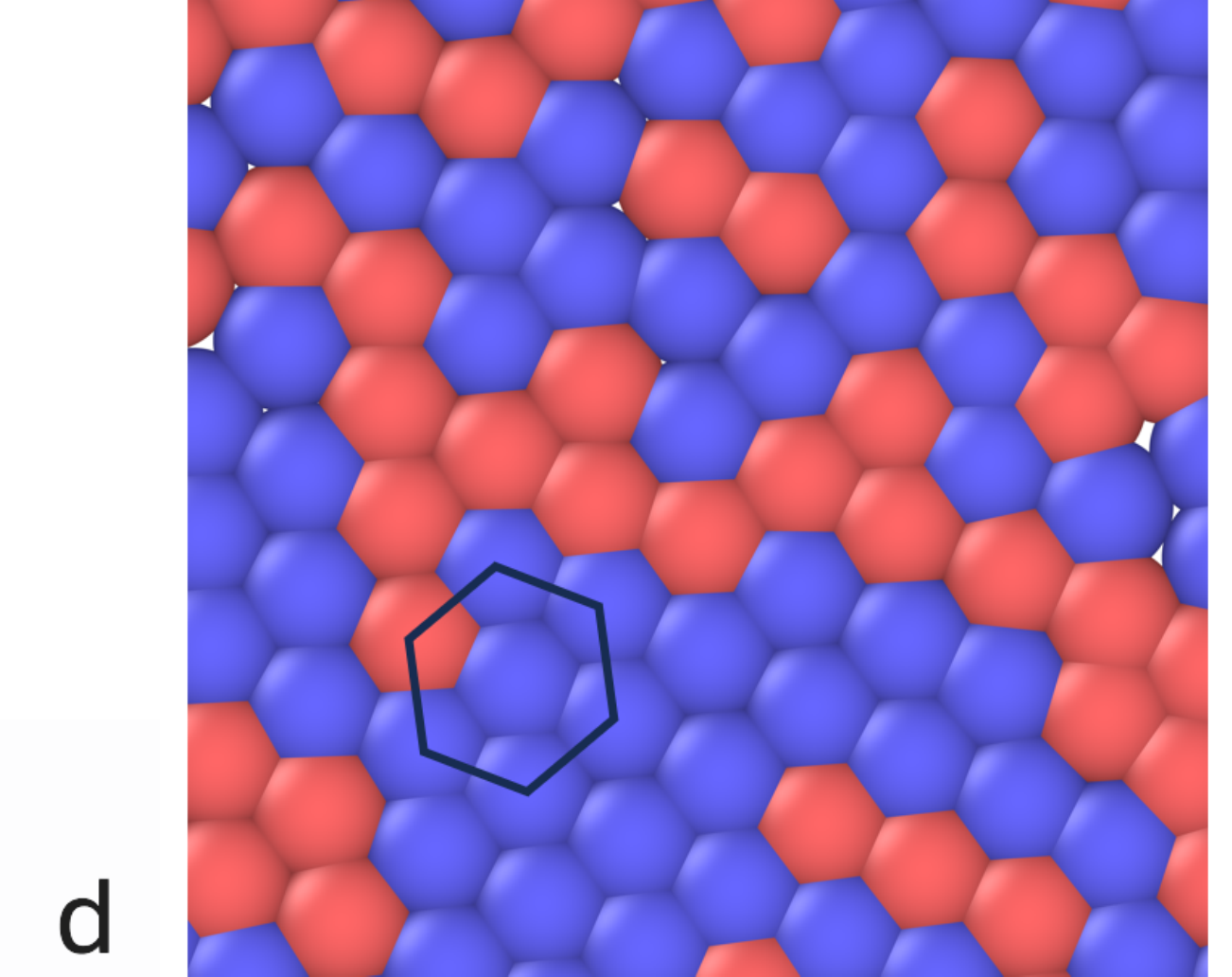}
\caption{ Figures (a), (b) and (c) illustrate the instantaneous snapshots of phase separating hot (red) and cold (blue) particles for density \(\rho^*=0.1\) and quench temperature \(T_h^*=25\) at different instances of time. Initially (\(t=5 \times 10^5\)), small circular cold nuclei form, which later coalesce into larger, elongated clusters (\(t=10^6,5 \times10^6\)). (b) Magnified view of a cold cluster showing hexagonal ordering of cold particles and trapping of hot particles within the cluster.}
\label{fig:snap}
\end{figure}

 We present here the results for \(T_h^*=25\), whereas results for \(T_h^*=40\) are given in Section 4 of the SM. The instantaneous snapshots of the phase separating binary mixture for  \(T_h^*=25\) at density \(\rho^*=0.1\) are given in Fig. \ref{fig:snap}(a-c). During the early stages post-quench, small, isolated nuclei of cold particles emerge within a sea of hot particles. These cold clusters initially maintain a nearly circular shape during their growth prior to reaching 1M time steps. Subsequently, these isolated cold clusters coalesce with each other to form larger elongated clusters, ultimately leading to phase separation. The cold clusters exhibit high-density crystalline order with the constituent particles arranged in a hexagonal lattice, as shown in Fig. \ref{fig:snap}(d), as they are compressed and stabilized by the kinetic pressure of the hot particles. Due to their crystalline order, cold clusters behave distinctly from liquid droplets undergoing phase separation \cite{puri2009kinetics,Binder2010,PhysRevE.85.050602}. While liquid droplets typically maintain circular shapes owing to dominant surface tension effects, the cold clusters here tend to form elongated or fractal morphologies due to the slower reorientation dynamics \cite{PhysRevLett.118.165701} of solid-like structures. We also note the enhanced trapping of hot particles in the phase separating cold clusters (Fig \ref{fig:snap}), particularly at the late stages. This characteristic of 2-TIPS at low dimensions \((d<3)\) \cite{D3SM00796K} is mainly due to the reduction of the cold-hot interface with decreasing dimension, thereby decreasing the number of pathways for hot particles to escape from the dense cold cluster. This mechanism of cluster coalescence is different from the phase separation kinetics of 2-TIPS at high density \cite{1c8b-hmxv}, where bicontinuous domains rich in either hot or cold particle grow progressively with time, akin to spinodal decomposition in passive systems.
\begin{figure}[hbt]
\centering
\includegraphics[scale=0.15]{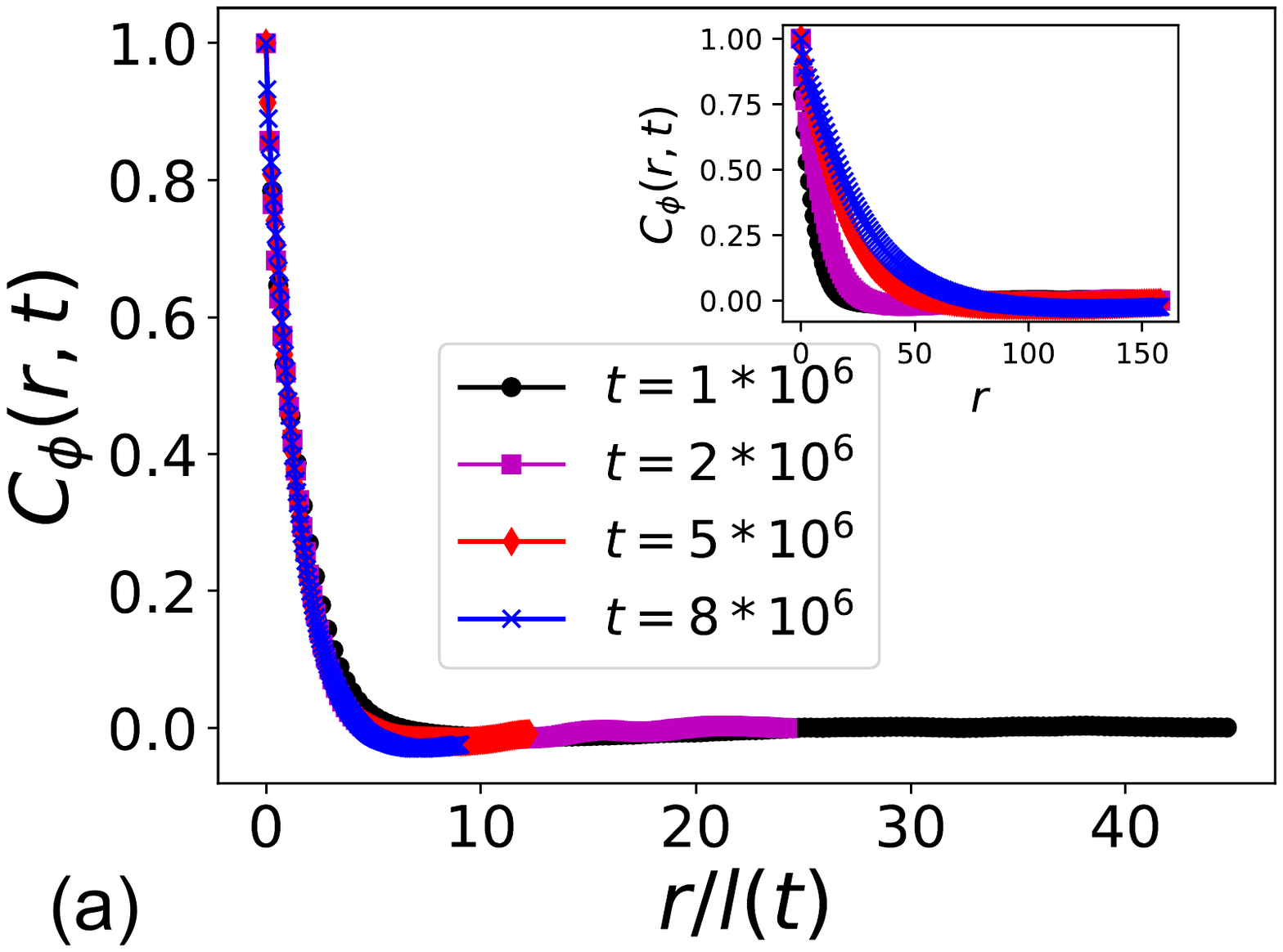} 
\includegraphics[scale=0.15]{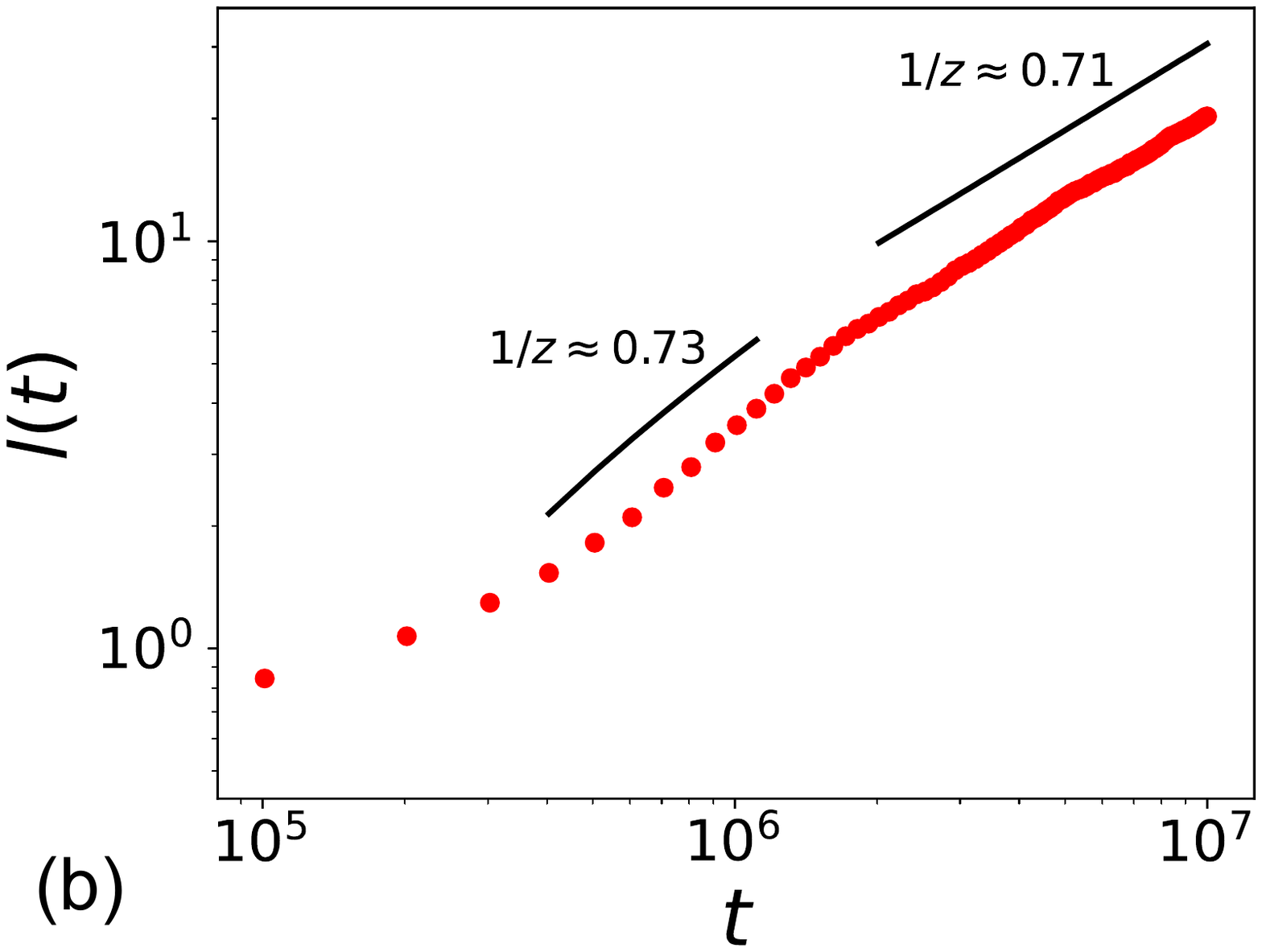} 
\caption{(a) Plot of the correlation function \(C_{\phi}(r,t)\) as function of rescaled distance \(r/l(t)\), where \(l(t)\) is the characteristic length at density \(\rho^*=0.1\) for \(T_h^*=25\) at different instants of time.  Inset illustrates plot of two-point spatial correlation function of \(\phi\), \(C_{\phi}(r,t)\) as function of distance between the points \(r\) at density \(\rho^*=0.1\) for \(T_h^*=25\) at different instants of time. The correlation function \(C_{\phi}(r,t)\) exhibits dynamic scaling. (b) Log-log plot of characteristic length \(l(t)\) versus time \(t\) shows algebraic growth of \(l(t)\). We see that the growth exponent \(1/z\) has a value close to 0.71 at late times.}
\label{fig:cor25}
\end{figure}

To quantify the domain morphologies of the phase-separating binary mixture, we first establish an order parameter field \(\phi(r,t)\) which can effectively distinguish the hot and cold domains, the details of which are elaborated in Section 2 of the SM. To obtain the average length of the cold domains, we calculate the two-point spatial correlation function of the order parameter field as defined below.
\begin{equation}
    C_{\phi}(r,t) = \langle\phi(\boldsymbol{r}_0,t)\phi(\boldsymbol{r}_0 +  \boldsymbol{r},t)\rangle -\langle\phi(\boldsymbol{r}_0,t)\rangle \langle\phi(\boldsymbol{r}_0 + \boldsymbol{r},t)\rangle
    \label{eq:cor}
\end{equation}
 where, $\langle \cdots \rangle$ denotes averaging over directions of $\boldsymbol{r}$ for a fixed $r=|\boldsymbol{r}|$, the reference point $\boldsymbol{r}_0$ as well as multiple independent realizations.
The characteristic length \(l(t)\), a measure of the domain size, is defined as the distance at which \( C_{\phi}(r,t)\) decays to half (0.5) of its value at $r=0$. Figure \ref{fig:cor25}(a) inset presents the plot of \( C_{\phi}(r,t)\) \emph{vs.} $r$ for \(T_h^*=25\), at different instants of time following the quench. We observe that decay of \( C_{\phi}(r,t)\) with \(r\) becomes slower over time, implying that the size of phase-separating domains increases as time progresses. After 1M timesteps post-quench, the phase separation kinetics enters a scaling regime. When plotted against the rescaled distance \(r/l(t)\), \( C_{\phi}(r,t)\) at different times collapse onto a single master curve as shown in Fig. \ref{fig:cor25}(a), indicating that the domain morphologies are statistically self-similar and differ solely in their characteristic length scale $l(t)$. In Fig. \ref{fig:cor25}(b), we present the plot of \(l(t)\) \emph{vs.} time $t$ for \(T_h^*=25\). At late times, we see that \(l(t)\) grows algebraically with time, typical of domain growth in kinetics, given by $l(t) \sim t^{1/z}$ ,where \(1/z\) is the growth exponent. Our calculations show that the value of the growth exponent at late times is approximately \(1/z \approx 0.71\). This value is significantly larger than the LS exponent of 1/3 observed at high density \cite{1c8b-hmxv}.

\begin{figure}[hbt]
\centering
\includegraphics[scale=0.14]{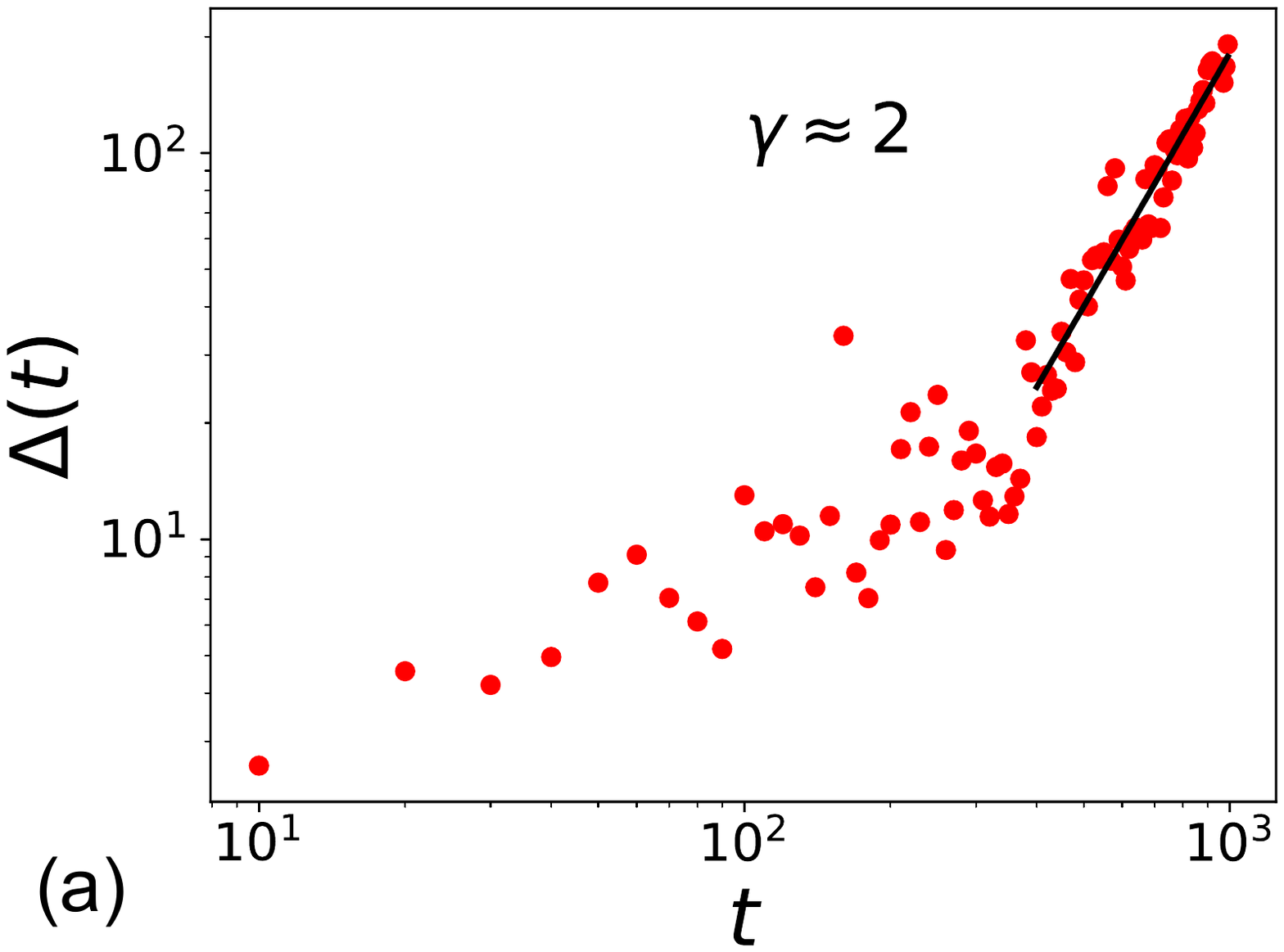} 
\includegraphics[scale=0.14]{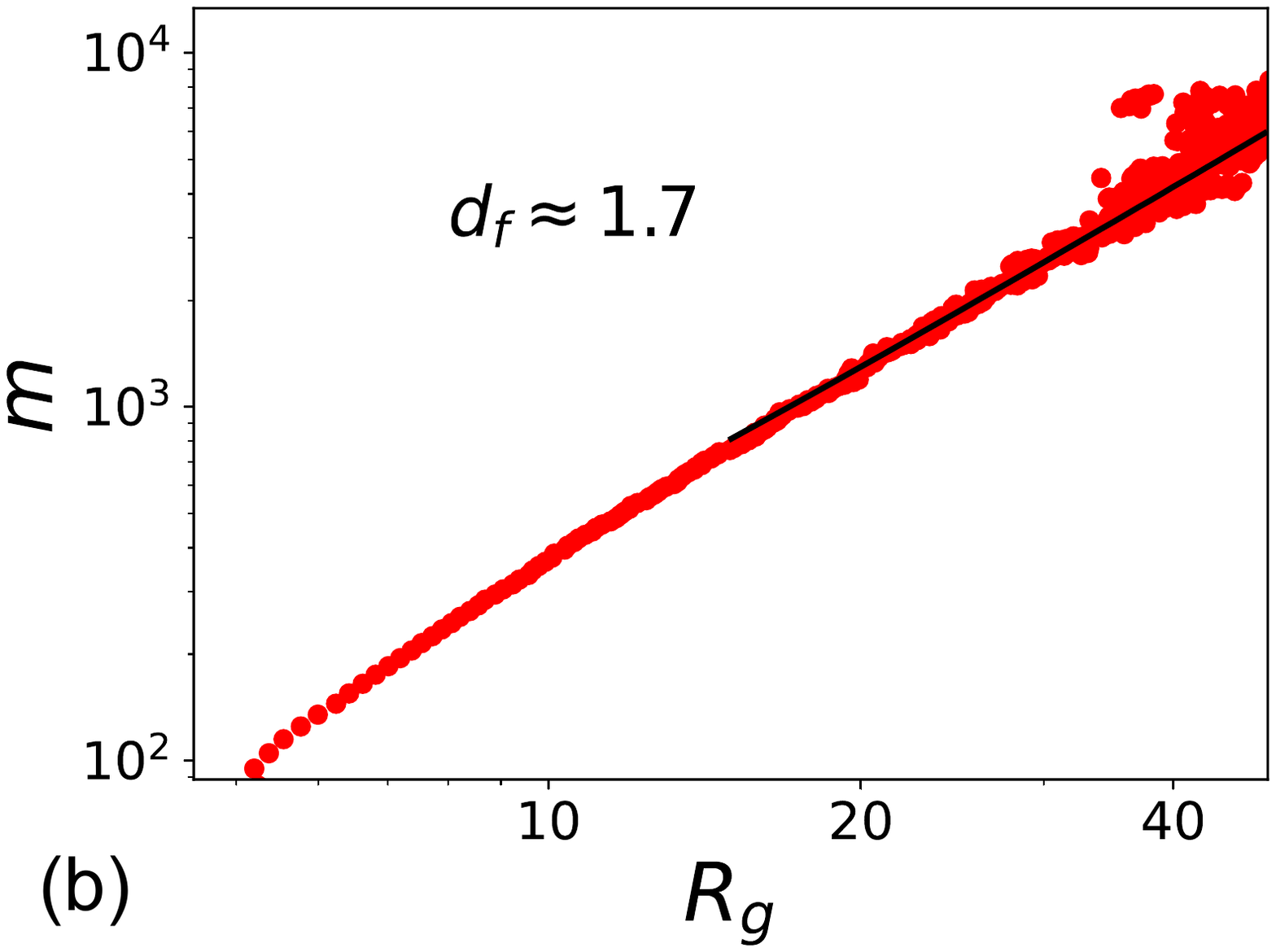}
\includegraphics[scale=0.14]{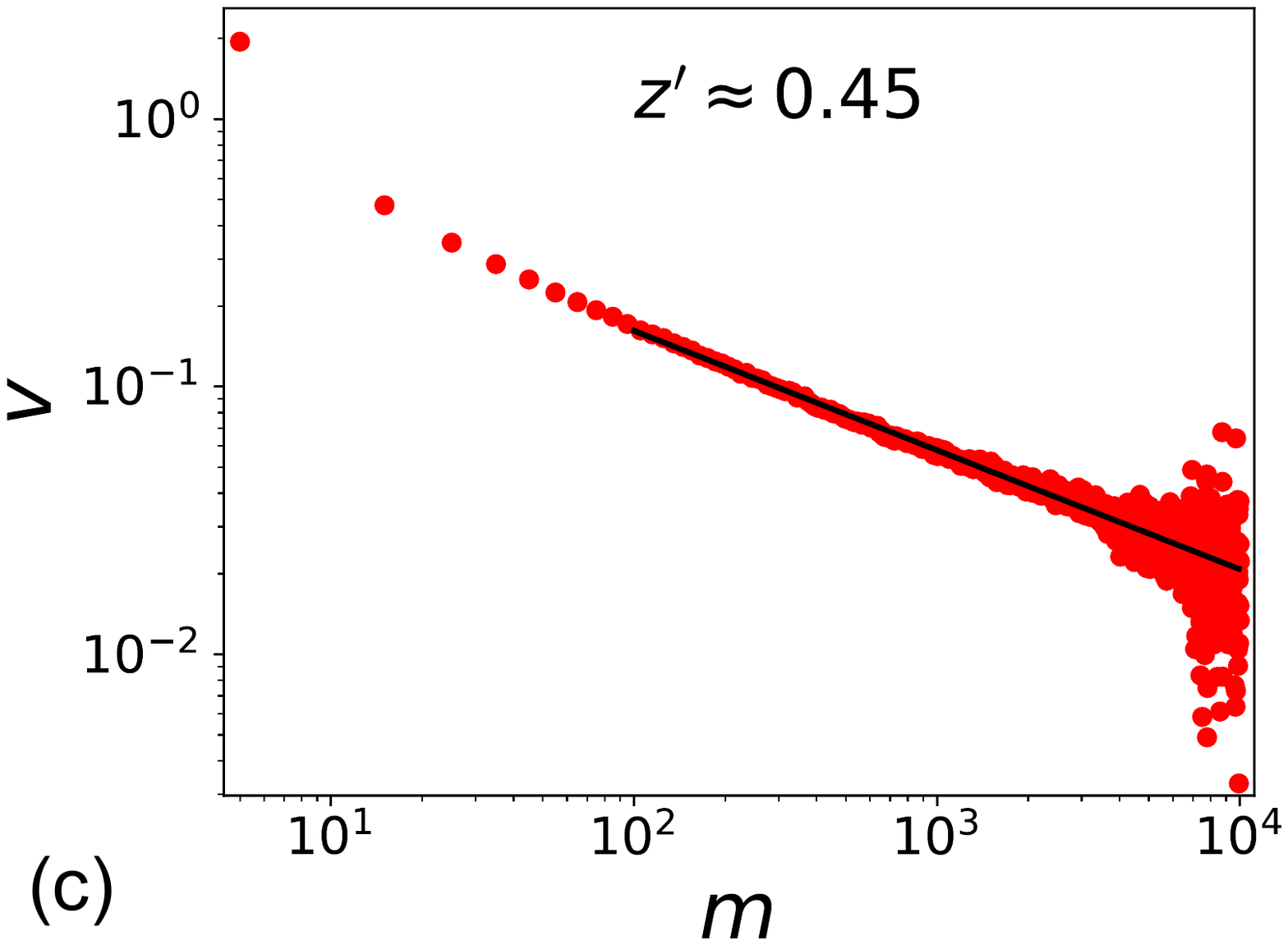} 
\caption{(a) The MSD \(\Delta(t)\) of cold clusters is plotted against time \(t\) on a log-log scale. The black curve represents the power law dependence of MSD on time with exponent \(\gamma \approx 2\), implying ballistic nature of cluster motion. (b) Log-log plot of average mass \(m\) of clusters versus average radius of gyration \(R_g\). The power law exponent associated (\(m \sim R_g^{d_f}\)) with it is \(d_f \approx 1.7\). (c) Log-log plot of average rms velocity of clusters \(v\) versus average mass \(m\). The associated power law decay exponent (\(v \sim m^{-z'}\)) is \(z' \approx   0.45\).}
\label{fig:expo}
\end{figure}

To investigate the origin of faster growth, we further analyze the coalescence mechanism by performing cluster analysis using the DBSCAN \cite{6814687} algorithm. Two particles are assigned to the same cluster if their separation is less than a cut-off distance \(r_c\). The value of \(r_c=1.2\sigma\) is obtained from the first peak of the radial distribution function of cold particles in the phase-separated steady state, evaluated from our previous studies \cite{chari2019scalar,D3SM00796K}. After identifying the cold clusters, we analyze the dynamics of the clusters prior to coalescence. However, the size of many of these cold clusters show large fluctuations with time. So, we carefully select 20 clusters which do not show large fluctuations in size during the time of their observations. We calculate the mean square displacement (MSD) $\Delta(t)$ of the center of mass (cm) of the selected clusters as, $\Delta(t)=\langle(\boldsymbol{r}_{cm}(t)-\boldsymbol{r}_{cm}(0))^2\rangle \sim t^\gamma$,
 where \(\boldsymbol{r}_{cm}\) is the position of cm of the cluster and \(\gamma\) denotes the scaling exponent associated with MSD. Figure \ref{fig:expo}(a) shows log-log plot of $\Delta(t)$ \emph{vs.} $t$, from which the value of the exponent \(\gamma\approx2\) is obtained. This indicates that the clusters exhibit ballistic motion preceding collision. The ballistic motion of cold clusters is counterintuitive, as one would anticipate a cold cluster immersed in a sea of hot particles to exhibit diffusive motion. However, a visual inspection of the cluster agglomeration process reveals that the isolated clusters move towards each other as if there is an effective attractive force between them. This may originate from the non-uniform forces exerted by the hot particles on the cold clusters, likely a consequence of their uneven concentration. Consequently, the motion of cold clusters is primarily ballistic rather than diffusive. In SM Movies (1 and 2), we show the dynamics of a few clusters, which clearly show the ballistic motion. 
 
Further, the cluster growth dynamics of ballistically moving clusters can be verified by theoretical results \cite{krapivsky2010kinetic,PhysRevLett.118.165701,PhysRevE.96.012105,D0SM01762K}, with details provided in Section 3 of the SM. The theory predicts the cluster mass growth exponent (\(m \sim t^{\beta}\)) as \(\beta = \frac{d_f}{d_f(1+z')-(d-1)},\)
where \(d_f\) and \(z'\) characterize the scaling of cluster mass with radius of gyration \(m \sim R_g^{d_f}\) and root mean square velocity \(v \sim m^{-z'}\), respectively. From Fig. \ref{fig:expo}(b) and (c), we estimate \(d_f \approx 1.7\) and \(z' \approx 0.45\), yielding \(\beta_{\text{theory}} \approx 1.16\) and a domain growth exponent \({1/z}_{\text{theory}} = \beta_{\text{theory}}/d_f \approx 0.68\), in close agreement with \(1/z \approx 0.71\) obtained from spatial correlation analysis. This consistency confirms ballistic agglomeration as the dominant phase separation mechanism in 2-TIPS at low density.

{\emph{Coarse-grained model}: We propose a coarse grained (CG) model to analyze the effective description of the system at a mesoscopic scale. In the mixture, the conserved order parameter fields corresponding to the density fields of the cold and hot species are denoted by $\psi_{\mathrm{c}}(\boldsymbol{r})$ and $\psi_{\mathrm{h}}(\boldsymbol{r})$, which are coupled to a velocity field $u(\boldsymbol{r})$ describing the local momentum flux due to both hot and cold species in the system. The dynamical equations at the mesoscopic level are given as follows,
\begin{equation}
    D_t\psi_{\mathrm{h}}=\nabla^2 \bigg{[} \alpha_\mathrm{h}\psi_\mathrm{h} +  \psi_\mathrm{h}^3 -  \nabla^2 \psi_\mathrm{h} -  \psi_\mathrm{c} \psi_\mathrm{h} + \frac{1}{2} \psi_\mathrm{c}^2 \bigg{]}
    \label{eq:dyn_h}
\end{equation}
\begin{equation}
    D_t\psi_{\mathrm{c}}=\nabla^2 \bigg{[}  \alpha_\mathrm{c}\psi_\mathrm{c} +  \psi_\mathrm{c}^3 -  \nabla^2 \psi_\mathrm{c} + \psi_\mathrm{h} \psi_\mathrm{c} - \frac{1}{2}\psi_\mathrm{h}^2 \bigg{]}
        \label{eq:dyn_c}
\end{equation}
\begin{equation}
    \rho D_t \boldsymbol{u} = \eta \nabla^2\boldsymbol{u}- \boldsymbol{\nabla}P + \boldsymbol{\nabla}\cdot\boldsymbol{\Sigma}
        \label{eq:dyn_vel}
\end{equation}
where, $D_t = \partial_t + \gamma(\boldsymbol{u}.\boldsymbol{\nabla})$,  $\gamma$ is the coupling strength.\\
All the terms on the right hand side of Eqs. \ref{eq:dyn_h} and \ref{eq:dyn_c} are introduced in our previous work \cite{1c8b-hmxv}. An additional equation for hydrodynamic velocity is introduced to incorporate the momentum of the droplets. In Eq. \ref{eq:dyn_vel} $\boldsymbol{\Sigma}$ denotes the stress tensor and has two contributions $\boldsymbol{\Sigma} =\boldsymbol{\Sigma}_p + \boldsymbol{\Sigma}_a$. $\boldsymbol{\Sigma}_p$ is the passive component of the stress tensor such that $\boldsymbol{\nabla} \cdot \boldsymbol{\Sigma}_p = -(\psi_h \boldsymbol{\nabla} \mu_h + \psi_c \boldsymbol{\nabla} \mu_c)$, where $\mu_c$ and $\mu_h$ are the chemical potential for the cold and hot species, respectively. The active component of stress is given by $\boldsymbol{\Sigma}_a = \chi_{cg} \bigg[(\partial_i\phi) (\partial_j\phi) - \frac{1}{2}\delta_{ij}(\boldsymbol{\nabla}\phi)^2  \bigg]$, where $\phi(\boldsymbol{r},t) = \psi_h(\boldsymbol{r},t) - \psi_c(\boldsymbol{r},t)$ describes the local phase separation order parameter between the cold and hot species. The coefficient $\chi_{cg}$ is the activity in the CG model and is defined as $\chi_{cg}=\alpha_h - \alpha_c$, which is a measure of the relative temperature difference between two species. It is important to note that the form of active stress and the results are unaffected under $\phi \to -\phi$ transformation. The active stress quantifies the excess interfacial stress generated by the temperature gradient between the interior of the cold domains and the surrounding hot medium for a phase separating droplet.  The details of Eqs.(\ref{eq:dyn_h}-\ref{eq:dyn_vel}) are described in Section 5 of the SM and the details of simulation of CG model is discussed Section 6 of the SM.

The average density of the cold and hot species is given by $\psi_{0,c/h}=\frac{1}{L^2}\sum_{\boldsymbol{r}}\psi_{c/h}(\boldsymbol{r})$. In our simulations, we keep hot and cold particles in equal proportions in the mixture i.e. $\psi_{0,c}=\psi_{0,h}=\psi_{0}$. $\psi_0 \approx 0$ signifies a densely packed system, where an increasingly lower value of $\psi_0$ implies lower densities. In this study we focus in the extreme low density regime by fixing $\psi_0 \in [-0.60,-0.70]$, which corresponds to the average density of cold and hot species in the range $\rho_{0,c/h}\in [0.15,0.20]$. Total density of the system is given by, $\rho_0 =\rho_{0,c}+\rho_{0,h}$.

In Fig. \ref{fig:cg_res}, time evolution of the mixture starting from a homogeneously mixed state is described through a series of snapshots of the $\phi$ -field. Starting from the homogeneous state, it takes some time before the domains of the cold particles ($\phi<0$) emerge as small droplets through nucleation (see MOVIE-3 in SM). Once the droplets appear, the size of the droplets grow rapidly with time mainly through coalescence (see MOVIE-4 in SM). To characterize the growth of the droplets, we calculated the two point correlation function $C_{\phi}(r,t)$ as defined in Eq. \ref{eq:cor}. The plot of $C_{\phi}(r,t)$ \emph{vs.} $r$ at different times is shown in the inset of Fig. \ref{fig:cg_res}(b). The slower decay of the correlation at successive times implies the growth of domains. The characteristics length, denoted by $l(t)$, is calculated as the $0.5$ crossing of the $C_{\phi}(r,t)$. The curve of $C_{\phi}(r)$ at different times collapse on a single master curve when plotted against the scaled distance $r/l(t)$ as shown in the main plot window of Fig. \ref{fig:cg_res}(b) (also, Fig.5(b) in SM). This implies the existence of a scaling regime where the domain morphology is independent of time and the domain growth follows $l(t)\sim t^{1/z}$ with an exponent $1/z$, where we found  $\frac{1}{z} \approx 0.70$ independent of density in range $\rho_0 \in [0.30,0.40]$, as shown in Fig. \ref{fig:cg_res}(c). We also observed that the growth exponent is independent of activity for the density in the given range (see Fig. 5(a) in SM). Thus, the growth law obtained from the CG model is consistent with that obtained from the MD simulations Fig. \ref{fig:cor25}.

In the scaling regime, domain growth proceeds primarily through droplet coalescence. We find that smaller droplets exhibit translational motion (see MOVIE-5 in SM), while larger droplets remain nearly immobile (see MOVIE-4 in SM). To characterize the dynamics of the smaller droplets, we calculate the MSD of the center of mass of the droplets, $\Delta (t)$. As shown in Fig. \ref{fig:cg_res}(d), $\Delta(t)$ varies as $\sim t^2$, which implies ballistic dynamics of the smaller droplets.
\begin{figure}[hbt]
    \centering
    \includegraphics[width=0.995\linewidth]{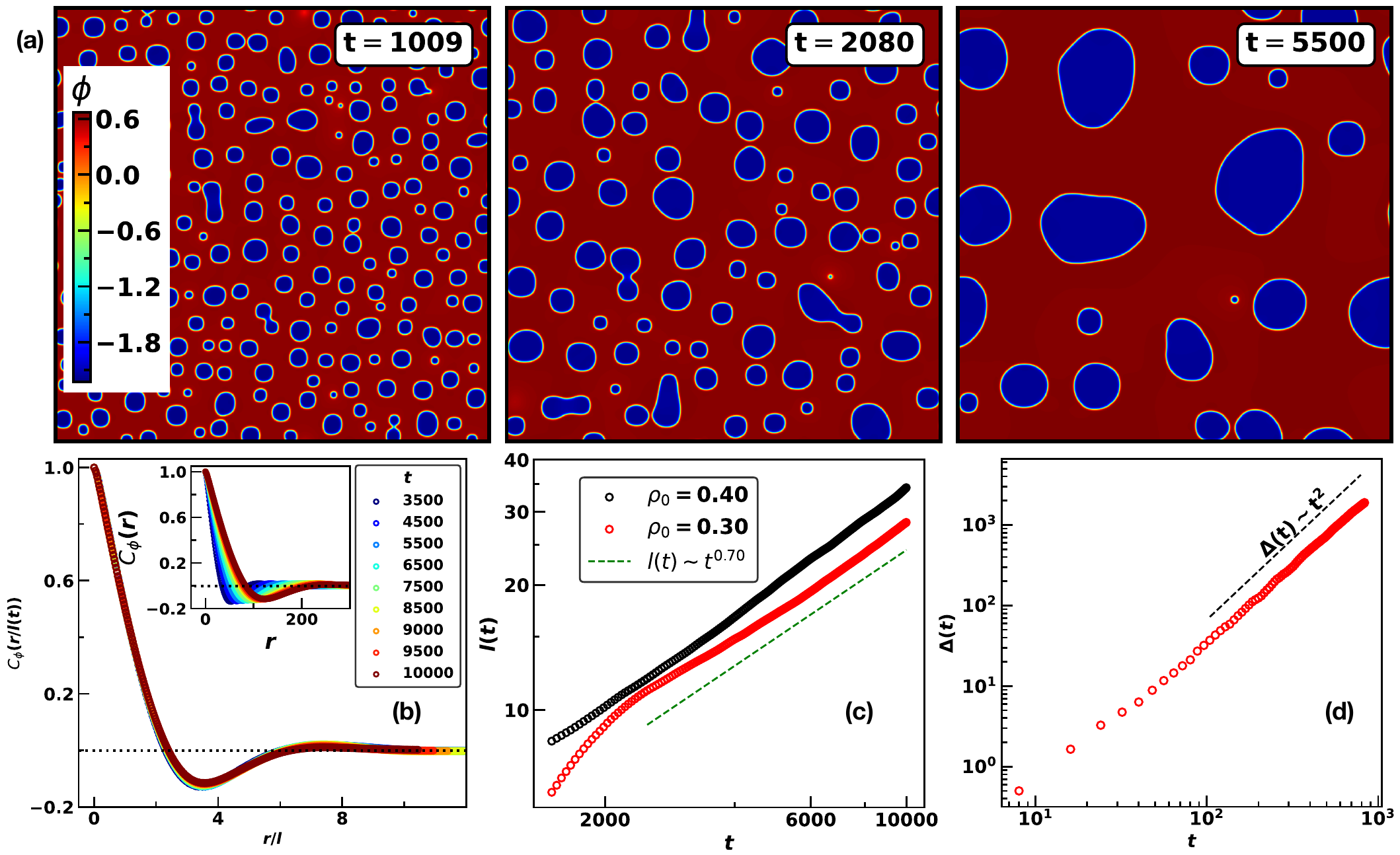}
    \caption{Panel (a) shows the emergence of phase separation in the mixture through series of snapshots at different times where the color shows the value of phase separation order parameter $\phi$ according to the colorbar. Panel (b) shows the plot of scaled correlation, $C(r/l)$ {\em vs.} $r/l$, at different times for $\rho_0 \approx 0.40$ and $\chi_{cg}=2.20$. (Inset) Plot of the unscaled correlation function, $C(r)$ {\em vs} $r$. Panel (c) shows the plot of time dependent characteristics length $l(t)$ {\em vs.} $t$ in log-log scale for $\chi_{cg}=2.20$ for $\rho_0 \approx 0.40$ and $0.30$, respectively. Panel (d) shows the plot of MSD of center of droplets in the scaling regime. Time $t=0$ in the plot is the reference time after which a droplet is tracked.}
    \label{fig:cg_res}
\end{figure}

In this Letter, we have examined the phase separation kinetics of 2-TIPS in the low-density regime using molecular dynamics (MD) simulations and developed a corresponding coarse-grained (CG) model to capture the system’s behavior at mesoscopic scales. The CG model couples the density order parameter fields to a velocity field that accounts for the local momentum flux arising from both hot and cold species.  Following the quench, as the cold particles begin to form domains, the resulting temperature contrast across domain boundaries generates an additional interfacial stress, whose magnitude scales with the temperature difference—i.e., the activity between the two species.

At low density, the hot and cold species phase separation proceeds via the nucleation of cold clusters embedded in a hot majority phase, whereas the same system at high density phase separates through the spinodal decomposition \cite{1c8b-hmxv}. After nucleation, these clusters undergo ballistic motion and grow through coalescence, leading to rapid coarsening. Our study shows that the domain size grows as $\approx t^{0.70}$ in both MD and CG methods. Furthermore, the growth exponent obtained from our study shows excellent agreement with the prediction from ballistic aggregation theory. The phase separation observed here is significantly different from that of EBMs in the dilute limit wherein the system reaches a metastable state and does not phase separate spontaneously. These results reveal a distinct kinetic regime from equilibrium mixtures and highlight the universal relevance of coalescence \cite{LEYVRAZ200395,krapivsky2010kinetic}, a process central to phenomena ranging from planet formation \cite{annurev:/content/journals/10.1146/annurev-earth-063016-020226,safronov1972evolution} and raindrop growth \cite{pruppacher1997microphysics} to colloidal aggregation \cite{lin1989universality,Royall_2021}.\\
The nonequilibrium phase separation kinetics in 2-TIPS is rich, showing non-trivial dependence on density. In the future, we would further extend our study of kinetics to include hot and cold sphero-cylinders, to see the influence of shape anisotropy on phase separation kinetics.

{\emph{Acknowledgment}}: P.K.M and N.V thank IRHPA grant from SERB (now ANRF) (No. IPA/2020/000034). P.S.M. thanks UGC for the research fellowship. S.M. thanks DST, SERB (INDIA), Projects No. CRG/2021/006945 and No. MTR/2021/000438 for financial support.

\bibliography{apssamp}

\end{document}